\documentclass{vietnam}

\usepackage{amsmath}
\usepackage{float}
\usepackage{multirow}
\usepackage[symbol]{footmisc}
\def\be{\begin{equation}}
\def\ee{\end{equation}}
\def\bea{\begin{eqnarray}}
\def\eea{\end{eqnarray}}

\newcommand{\Photo}{\includegraphics[height=35mm]{mypicture}}
\begin{document}
\vspace*{2cm}
\title{Develop monitors for MW-power proton beam at J-PARC extraction beamline for neutrino experiments }

\author{ S. Cao (IFIRSE, ICISE)\footnote[1]{Contact: cvson@ifirse.icise.vn, presented at PASCOS 2024, ICISE, Quy Nhon, Vietnam},  M. Friend (IPNS, KEK)   }



\maketitle

\abstracts{
Intense and well-characterized neutrino sources, coupled with large and high-performance detectors, are essential for elucidating the unknowns in leptonic mixing through neutrino oscillation measurements. The J-PARC accelerator and neutrino extraction beamline have recently undergone upgrades, successfully delivering an 800-kW beam. We discuss the development, recent findings, and outstanding challenges associated with a non-destructive beam-induced fluorescence monitor and optical-fiber-based beam loss monitors.}

\section{The need of intense and well-controlled accelerator-based neutrino beam}
Neutrinos, once thought to be massless, are now known to have a tiny but finite mass, revolutionizing the Standard Model of particle physics.  A paradigm based on a $3\times 3$ unitary mixing matrix known as the PMNS matrix has been developed and so far fit well with the global neutrino data landscape~\cite{ParticleDataGroup:2024cfk}. However, open questions still remain including, but not limited to, the existence and actual magnitude of CP violation, ordering of neutrino mass, and completeness of PMNS matrix itself. Apparently, more sensitive neutrino oscillation data is mandatory, necessitating intense and well-controlled accelerator-based neutrino beams for addressing these questions. 

\emph{--- J-PARC extraction neutrino beamline}, detailed in Ref.~\cite{Abe:2011ks}, have been built to provide intense and well-controlled neutrino beam for long-baseline neutrino oscillation experiments. An 800~kW operation has been achieved in 2024 at $2.3 \times 10^{14}$  protons-per-pulse (ppp) with a cycle of 1.36~s. By 2026, J-PARC aims to reach 1.3 MW with a beam intensity of $3.2 \times 10^{14}$ ppp  at a 1.16~s cycle. For each cycle, eight circulating proton bunches are extracted in a single turn to the neutrino beamline by kicker magnets. The primary 238 m-length neutrino beamline, composed of 21 normal conducting magnets and 14 doublets of superconducting combined function magnets, accept, transport, and bend beam toward the neutrino detectors. A secondary beamline, situated downstream of the primary one, houses a target station equipped with 91.4 cm-length graphite rod and a trio of magnetic horns, a decay volumne, beam dump and muon monitor for the production, filtration, and monitoring of the neutrino beam. To achieve MW-level power beam~\cite{T2K:2019eao}, it is essential to ensure equipment robustness against high intensity, tolerability of beam loss, management of radioactive waste, and precise, continuous monitoring of the beam. 

\emph{--- Critical beam information for continuous operation:} Numerous beam monitors~\cite{Abe:2011ks,Friend:2019fuq} have been installed along the neutrino beamline to guarantee safe operation, accurately track the beam trajectory, assess the state of beam equipment, and evaluate parameters relevant to predicting neutrino production. Beam losses, which may arise from factors such as beam instability, transport efficacy, mis-steering, malfunctioning components, and vacuum breaks in the beam pipe, are monitored using the beam loss monitor (BLM) system. The loss of beam can lead to serious consequences, including damage to beamline equipment, quenching of superconducting magnets, and radioactivity of components, raising concerns regarding human safety and environmental protection. In J-PARC, the BLM system is incorporated into the Machine Protection System to abort the beam whenever BLM signals exceed established thresholds. The beam intensity is measured with $\sim 2.7\%$ uncertainty using five current transformers, each consisting of a 50-turn toroidal coil wound around a ferromagnetic core. The beam center position and angle at several beam segments are measured nondestructively using  21 of electrostatic monitors (ESMs), each composing of four segmented cylindrical electrodes symmetrically arranged around the beam trajectory, allowing for precise measurements exceeding 500~$\mu m$. The transverse profiles of the proton beam are measured using 19 Ti-foil (5~$\mu m$-thick) segmented secondary emission monitors (SSEM) and one optical transition radiation (OTR) sensor positioned upstream of the target. Measurement with SSEM in bunch-by-bunch basics can achieve 0.07~mm and 0.2~mm uncertainty for the beam center position and beam width respectively.  Each SSEM that intercepts the beam causes a 0.005\% loss of beam power. Excluding the most downstream SSEM, other 18 SSEMs are utilized only periodically for beam tuning and studying.
\section{Development of non-destructive beam profile monitors}\label{sec:bif}
The continuous operation of beam profile monitors presents challenges~\cite{Friend:2019qds} due to concerns regarding beam loss, soared radiation, and degradation of the monitors. A non-destructive beam profile monitor currently under development at the J-PARC neutrino beamline may offer a solution.

\emph{--- Concept of beam-induced fluorescence (BIF) monitor}: The development of the BIF and the initial observation of signals are detailed in Ref.~\cite{Friend:2020lpp} and references herein. In principle, the transverse profile of fluorescence resulting from proton interactions with injected gas matches with that of the proton beam. Nitrogen gas is selected for its large cross section with proton, short-lifetime ( $\sim 60$~ns) and visible fluorescence (dominated line of 391~nm wavelength), along with its technical compatibility with existing gas-pumping system in the beamline. A remotely controlled system has been developed to inject gas through a series of valves, achieving a localized gas pressure of approximately $10^{-2}-10^{-3}$ Pa in the measurement region. Gas must be evacuated using an ion pump to achieve a pressure of approximately $10^{-5}$~Pa between proton spills. An interlock system is designed to close a remotely-controlled pneumatic valve when pressure exceeds a predetermined threshold. Gas is presently injected in a singular spill as required. Fluorescence capture is done through  view ports mounted on the beam pipe, necessitating two reading systems for horizontal and vertical profiles. 
\begin{figure}
    \centering
\includegraphics[width=0.4\linewidth]{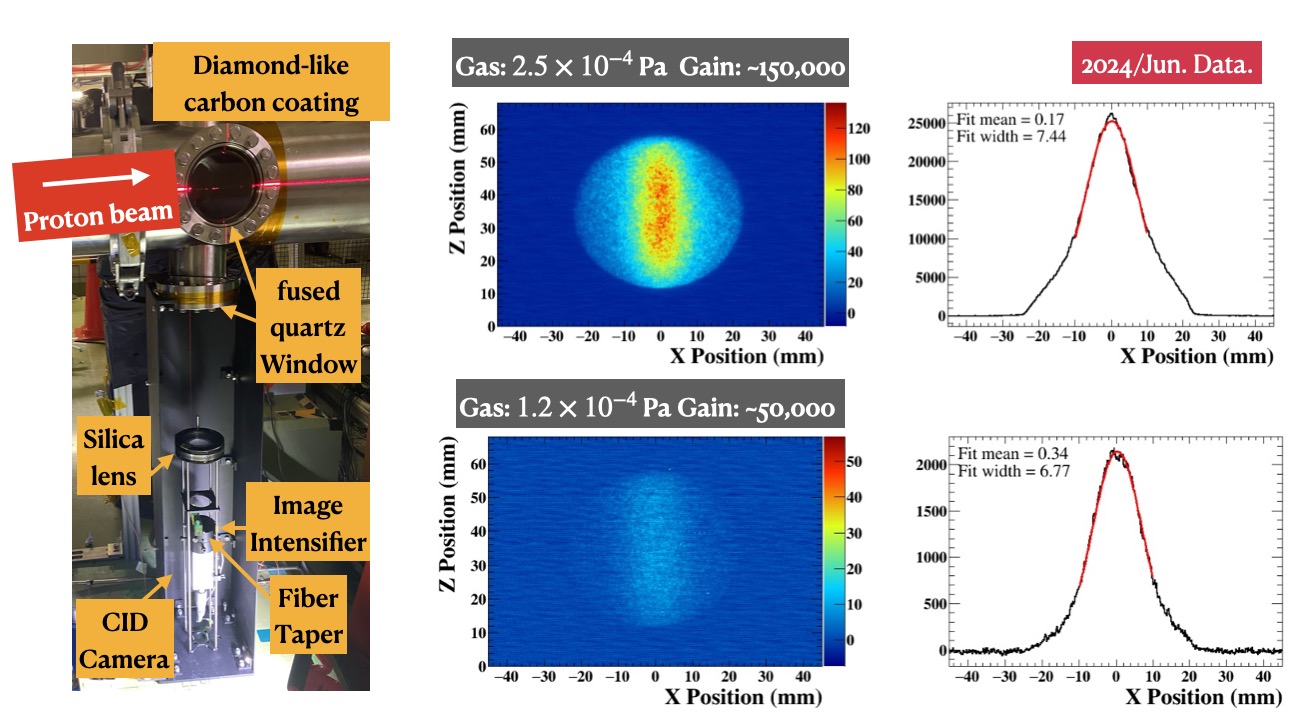}
    \includegraphics[width=0.43\linewidth]{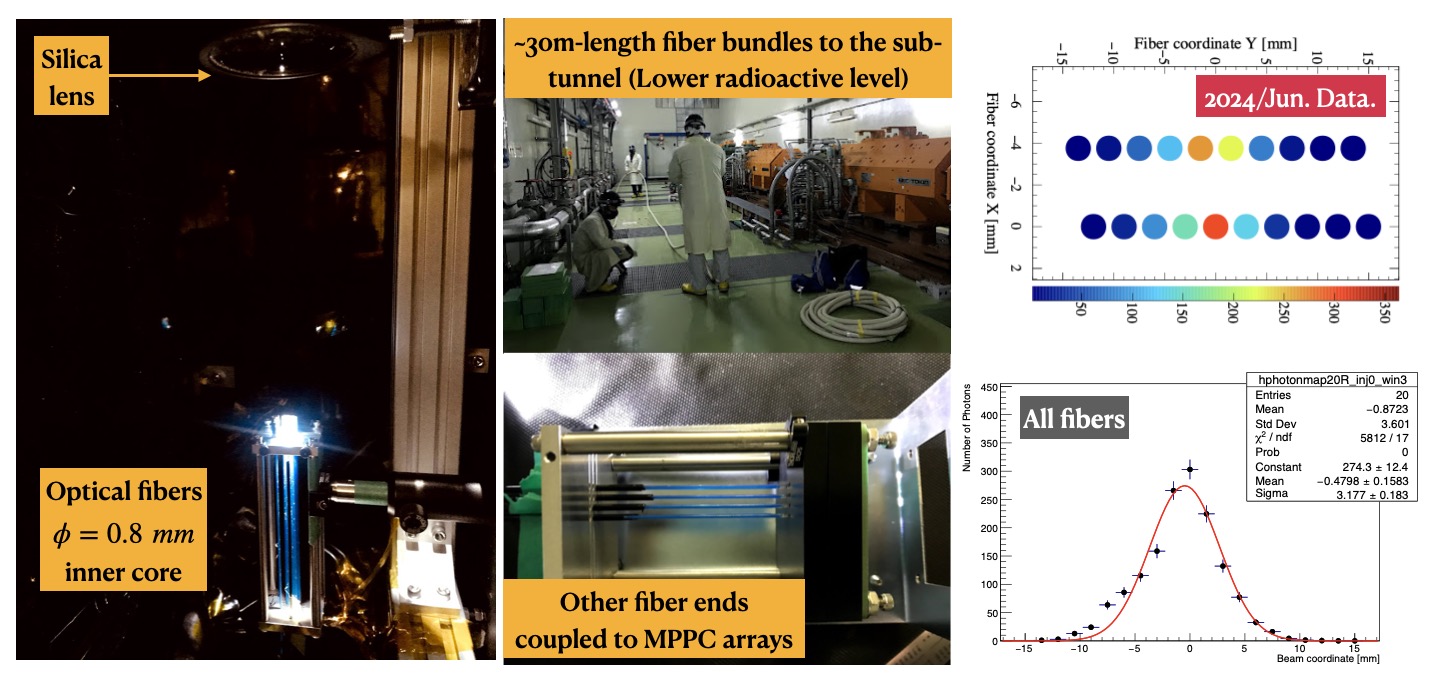}
    \caption{Left: Optical readout with CID camera and image intensifier. Right: Optical readout with MPPC array and optical fibers. The measured profiles are with the gas injection test conducted at 800~kW proton beam.}
    \label{fig:bif_prof}
\end{figure}

\emph{--- Optical readout with charge injection device (CID) camera and image intensifier:}
Radiation-hard and low-noise CID camera is selected for capturing the proton beam's transverse profile in x-axis. Two 75~mm-diameter UV fused silica plano-convex lens  are staged to focus the fluorescence light onto the micro-channel plate-based image intensifier, which is coupled to CID camera via a fiber taper. For a beam test in 2024, a new fast-gateable image intensifier with higher gain than the previous setup to detect fainter light is used. Also, it couples with a new photocathode with selective wavelength band for better signal-to-background ratio. The setup and measured profiles are shown in Fig.~\ref{fig:bif_prof}. Preliminary analysis indicates a strong correlation between beam position measurements obtained from the camera and those from nearby ESM monitors. The beam width recorded by the camera is slightly broader than that of the neighboring WSEM. Further investigation is ongoing.
\begin{figure}
    \centering
  \includegraphics[width=0.6\linewidth]{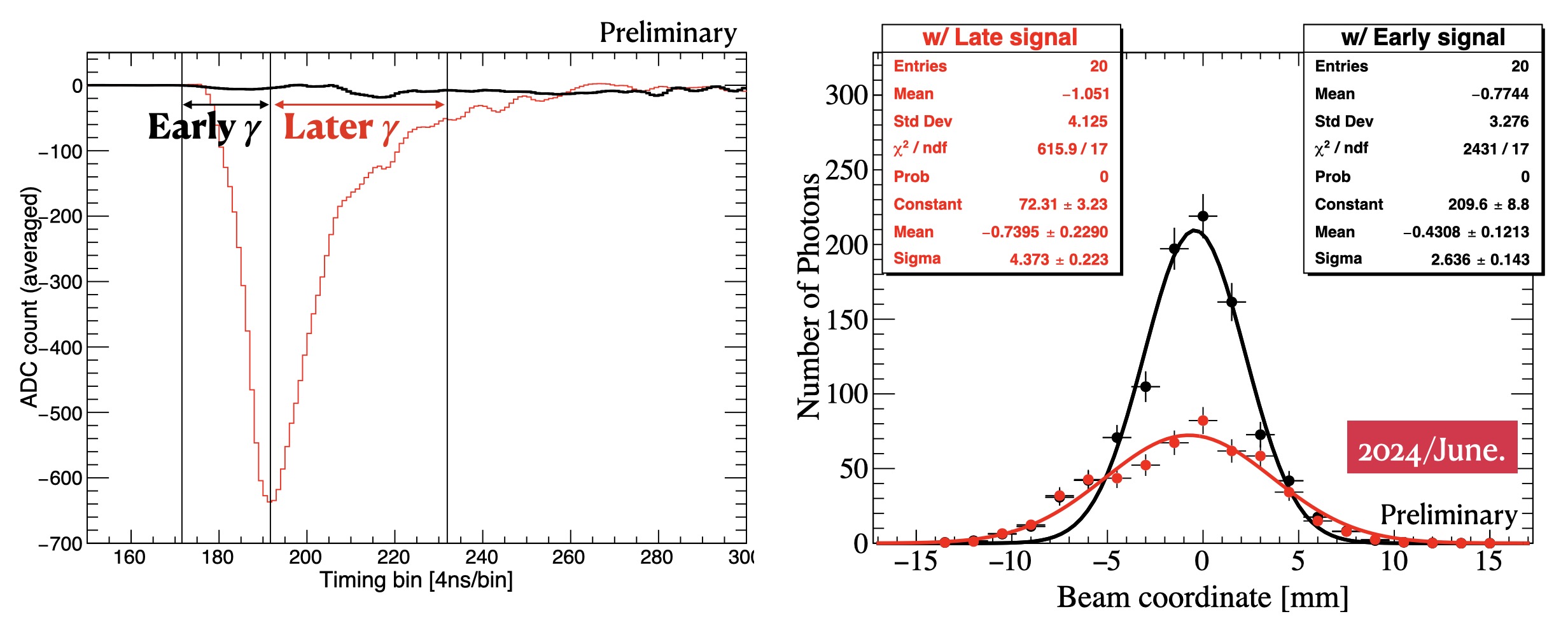}
 \caption{Reconstructed beam profile are different with early photon and later photon implying the possible space-charge effect.}
    \label{fig:bifspace}
\end{figure}

\emph{--- Optical readout with Multi-pixel photon counter (MPPC) array and optical fibers} A fast-timing response optical readout is desired for studying the potential impact of so-called space-charge effect on the beam profile reconstruction, as ionized gas molecules can be drifted by the strong transverse beam-induced field. The MPPC array is selected due to its significant attributes, including high gain, fast response, portability, magnetic resistance, and cost efficiency. However, MPPC lacks radiation hardness; therefore, optical fiber has been utilized to transport the fluorescence light from the high-radiation-level beamline to a lower radiation area for coupling with the MPPC. A pair of plano-convex silica lenses is employed to concentrate the fluorescence light from the beamline onto the array of optical fiber ends. Fig.~\ref{fig:bif_prof} illustrates the setup and reconstructed profile utilizing an 800 kW proton beam from the gas injection test conducted in 2024. A preliminary result indicates stability in the reconstructed profile. It is noteworthy that the reconstructed profile with early light is narrower than that with later light. This may suggest the presence of a space charge effect due to $\sim 3\times 10^{13}$ protons per bunch. Additional analysis is required prior to drawing a definitive conclusion on this matter.  

\emph{---Remaining challenges:} Optimal configurations of gas injection system are under investigation for both adequate fluorescence generation and tolerance to the beamline's vacuum level.  Accurate calibration of both optical readout systems must be conducted to account for various optical and electronic effects.  We must continue to examine the space-charge effect on the reconstructed profiles. Verification with adjacent beam monitors and modeled beam optics is crucial to ascertain the accuracy of this innovative beam profile monitor. 

\section{Development of optical fiber-based beam loss monitors }\label{sec:oblm}
\begin{figure}
    \centering
    \includegraphics[width=0.7\linewidth]{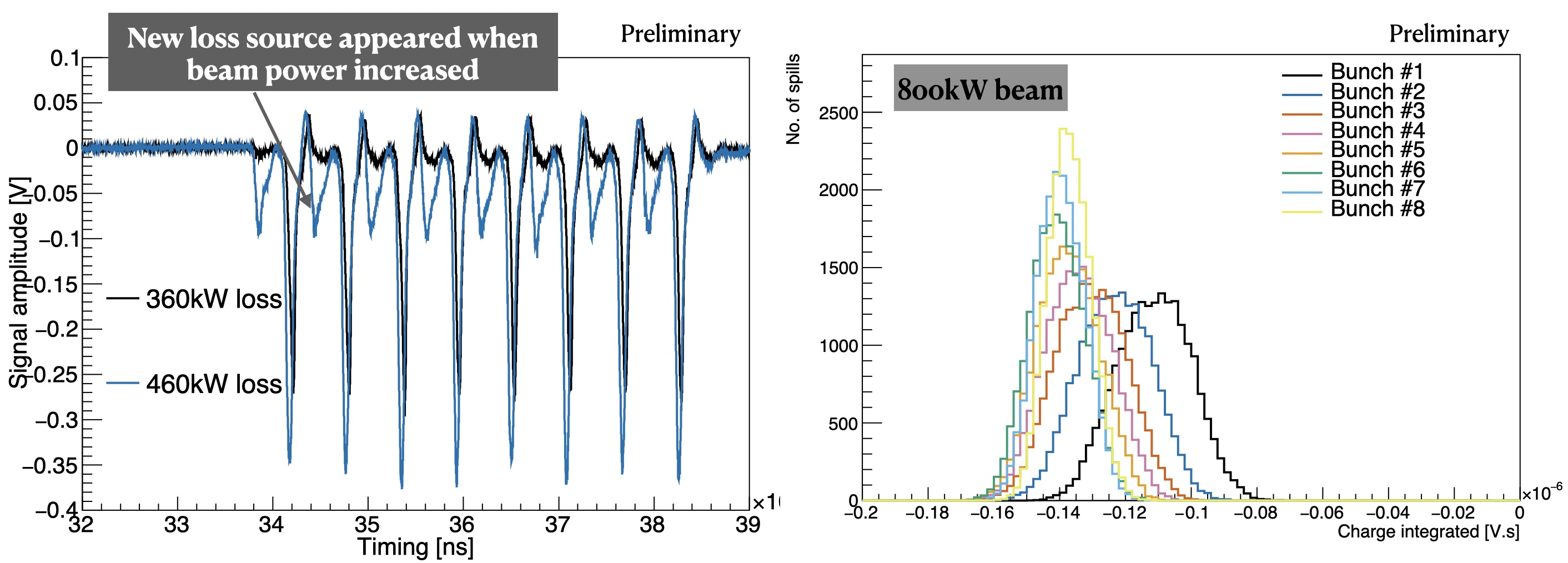}
    \caption{Left: O-BLM signal at different beam powers. Right: Beam loss measured by O-BLM in bunch-by-bunch basics.}
    \label{fig:oblm_sig}
\end{figure}

The R\& D of optical-fibers beam loss monitors (O-BLM)~\cite{Cao:2020bpn} as an alternative to gas-based BLMs is driven by the need for bunch-by-bunch monitoring of beam loss, the potential for identifying loss locations, flexibility in triggering, sampling timing window, and data acquisition, as well as cost-effectiveness with minimal number of hardware and electronic channels. Secondary charged particles generated from the interaction of lost protons with beamline materials produce Cherenkov light as they traverse the optical fiber. In an optimal configuration for selected radiation-hard silica-core optical fiber, where proton passage opposes the trajectory of collected Cherenkov light, a loss location can be determined with a resolution of few 10~cm if the photosensor exhibits a ns rise time. The limitation for loss source separation in the J-PARC neutrino beamline is caused by a relatively broad bunch width of approximately 13~ns, which results in an overlap of the Cherenkov signal when the two loss sources are in proximity. A basic simulation yields a resolution of around 7~m, supposing the two loss sources are of comparable magnitude. In the 2024 beam test, we have instrumented approximately 90~m of beamline, specifically covering the whole preparation and final focusing sections of the primary beamline. The losses in the arc section with superconducting magnets are insignificant, given that optical fiber is challenging to place near the beam center. The signal is captured using metal-package PMTs and recorded with a 500 MHz sampling oscilloscope. The waveform of a single spill with varying beam power is depicted in Fig.~\ref{fig:oblm_sig}. It clearly illustrates the eight-bunch structure of the loss. The higher signal peak in each bunch is the primary cause of loss, possibly resulting from beam scattering directly on the beam window at the downstream end of the beamline. A new loss source, seen as the smaller peak for each bunch, emerges with an increase in beam power. The new source of loss occurs earlier in time as it is nearer to the fiber end connected to the PMT. The O-BLM, characterized by fast response and a distinct bunch structure, facilitates the monitoring of beam loss on a bunch basis, hence aiding in the identification of any issues related to the beam bunch structure. 

\emph{---Remaining challenges:} An unbiased understanding of the beam's longitudinal proton profile is crucial, as the loss signal is convoluted with it. The loss signal exhibits a significant dynamic range, approximately 300-fold variation with beam intensity and 20-fold variation between continuous operation and beam tuning at the same power, which presents challenges for recording with optical fiber. Attenuation in long optical fibers and electrical cables presents a challenge for precise timing profile measurements.  The radiation hardness of optical fibers and photomultiplier tubes (PMTs) requires evaluation for long-term operation.
\section{Summary and prospects}
 The beam-induced fluorescence monitor has been developed and upgraded, successfully reconstructing the proton beam profile with an 800 kW proton beam. The optical-fiber beam loss monitor has, for the first time, instrumented 90~m of the beam line using two long optical fibers couplings with fast-response PMTs. This setup successfully monitors the beam on a bunch-by-bunch basis and provides valuable information for identifying the sources of beam loss. Both present remaining challenges for completion and implementation as novel proton beam monitors. 
\section*{Acknowledgement}
S. Cao would like to thank Neutrino group, IPNS, KEK for their support. The research of S. Cao is funded by the National Foundation for Science and Technology Development (NAFOSTED) of Vietnam under Grant No. 103.99-2023.144.

\end{document}